\documentstyle[12pt]{article}
\def\nabstar#1{\nabla\kern-0.5pt\smash{\raise 4.5pt\hbox{$\ast$}}
               \kern-4.5pt_{#1}}

\def\drvstar#1{\partial\kern-0.5pt\smash{\raise 4.5pt\hbox{$\ast$}}
               \kern-5.0pt_{#1}}

\def\newline{\relax\ifhmode\null\hfil\break\else\nonhmodeerr@\newline\fi}
\def\frac#1#2{{#1\over#2}}
\def\text#1{{\hbox{\rm #1}}}
\def\flushpar{{\par \noindent}}

\newcommand{\beq}{\begin{equation}}
\newcommand{\eeq}{\end{equation}}
\newcommand{\bea}{\begin{eqnarray}}
\newcommand{\eea}{\end{eqnarray}}
\def\Id{ \mbox{1\hspace{-1.2mm}I} }
\def\BE{\begin{equation}}
\def\EE{\end{equation}}
\def\BA{\begin{eqnarray}}
\def\EA{\end{eqnarray}}
\def\BAN{\begin{eqnarray*}}
\def\EAN{\end{eqnarray*}}

\def\gm5{\gamma^5}

\begin{document}
\thispagestyle{empty}
\begin{flushright}
NTUTH-99-828 \\
August 1999
\end{flushright}
\bigskip\bigskip\bigskip
\vskip 2.5truecm
\begin{center}
{\LARGE {A construction of chiral fermion action }}
%\vskip 7pt
\end{center}
\vskip 1.0truecm
\centerline{Ting-Wai Chiu}
\vskip5mm
\centerline{Department of Physics, National Taiwan University}
\centerline{Taipei, Taiwan 106, Republic of China.}
\centerline{\it E-mail : twchiu@phys.ntu.edu.tw}
\vskip 2cm
\bigskip \nopagebreak \begin{abstract}
\noindent

According to the necessary requirements for a chirally symmetric Dirac
operator, we present a systematic construction of such operators.
We formulate a criterion for the hermitian operator which enters the
construction such that the doubled modes are decoupled even at finite
lattice spacing.


\end{abstract}
\vskip 1.5cm

\newpage\setcounter{page}1

Recently we have discussed the necessary requirements \cite{twc98:9a}
for a chirally symmetric Dirac operator $ D_c $ to enter
the topologically invariant transformation
$ D = D_c ( \Id + R D_c )^{-1} $ \cite{twc98:6a} such that the resulting
Dirac operator $ D $\footnote{ Note that $ D $ satisfies the Ginsparg-Wilson
relation \cite{gwr} automatically. }
could reproduce the continuum physics,
where $ R $ is a hermitian operator which is local in the position
space and trivial in the Dirac space. These constraints for
$ D_c $ are :
\begin{description}
\item[(a)] $ D_c $ agrees with $ \gamma_\mu ( \partial_\mu + i A_\mu ) $
             in the classical continuum limit.
\item[(b)] $ D_c $ is nonlocal.
\item[(c)] $ D_c $ is free of species doubling.
\item[(d)] $ D_c $ is well defined in topologically trivial background
             gauge field.
\item[(e)] $ D_c $ has zero modes as well as simple poles ( or equivalently,
             $ V $ has real eigenvalue pairs $ \pm 1 $ ) in topologically
             non-trivial background gauge fields. Furthermore, the zero
             modes of $ D_c $ satisfy the Atiyah-Singer index theorem
             for any prescribed smooth gauge background.
\end{description}

In this paper, we attempt to construct such chirally symmetric Dirac
operators $ D_c $ which also satisfies the hermiticity condition
\bea
\gamma_5 D_c \gamma_5 = D_c^{\dagger} \ .
\label{eq:hermit}
\eea
The hermiticity condition and the chiral symmetry of $ D_c $ implies that
$ D_c $ is antihermitian, thus there exists one to one correspondence
between $ D_c $ and a unitary operator $ V $,
\bea
D_c = (\Id + V )(\Id - V )^{-1}, \hspace{4mm}
V = (D_c - \Id)( D_c + \Id)^{-1}.
\label{eq:VDc}
\eea
where $ V $ also satisfies the hermiticity condition
$ \gamma_5 V \gamma_5 = V^{\dagger} $.
Then the unitary operator $ V $ can be expressed in terms of
a hermitian operator $ h $,
\bea
\label{eq:V5h}
V = \gamma_5 h = \left( \begin{array}{cc}
                         \Id  &    0    \\
                           0  & -\Id
                        \end{array}      \right)
                  \left( \begin{array}{cc}
                          h_1            &  h_2    \\
                          h_2^{\dagger}  &  h_3
                        \end{array}      \right)
               =  \left( \begin{array}{cc}
                          h_1            &  h_2    \\
                         -h_2^{\dagger}  & -h_3
                        \end{array}      \right)
\eea
where $ h_1^{\dagger} = h_1 $ and $ h_3^{\dagger} = h_3 $.
Using the unitarity condition $ V^{\dagger} V = \Id $,
we have $ h^2 = \Id $,
\bea
\label{eq:h^2}
h^2 = \left( \begin{array}{cc}
     h_1^2 + h_2 h_2^{\dagger}              &   h_1 h_2 + h_2 h_3    \\
     h_2^{\dagger} h_1 + h_3 h_2^{\dagger}  &   h_2^{\dagger} h_2 + h_3^2
            \end{array}      \right)
    = \left( \begin{array}{cc}
              \Id          &  0    \\
                0          &  \Id
             \end{array}      \right)
\eea
Then we obtain \cite{twc98:10b}
\bea
D_c
\label{eq:Dc}
\equiv \left[ \begin{array}{cc}
                     0          &   D_R  \\
                     D_L        &   0
              \end{array}                           \right]
=  \left[ \begin{array}{cc}
                     0                 &  (\Id - h_1 )^{-1} h_2  \\
     -h_2^{-1} (\Id + h_1)             &   0
              \end{array}                           \right]
\eea
where $ D_L = - D_R^{\dagger} $.
The general solution to Eq. (\ref{eq:h^2}) can be written as
\bea
\label{eq:h1}
h_1 &=& \pm \frac{1}{ \sqrt{ \Id + b b^{\dagger} } } \\
\label{eq:h2}
h_2 &=& \frac{1}{ \sqrt{ \Id + b b^{\dagger} } } \ b \ e^{ i \theta }
\eea
where $ b $ is any operator with phase factor $ \exp( i \theta ) $.
In the following we will restrict $ \theta $ to zero,
and also pick the minus sign for $ h_1 $.
Then the general solution for $ D_c $ can be written in the following
form
\bea
\label{eq:DL}
D_L &=& b^{-1} \left[ \Id - \sqrt{\Id + b b^{\dagger}} \ \right] \\
\label{eq:DR}
D_R &=& \left[ \Id + \sqrt{\Id + b b^{\dagger}} \ \right]^{-1}  b
\eea
Due to the presence of the square root in (\ref{eq:DL}) and (\ref{eq:DR}),
$ D_c $ is nonlocal for nontrivial $ b $, thus the requirement (b) is
satisfied. Then the Nielson-Ninomiya no-go theorem \cite{no-go} is
circumvented.
Hence, we can construct $ D_c $ to be free of species doubling. If $ D_c $
is topologically proper [ i.e., satisfying the constraint (e) ], then we
can use the topologically invariant transformation
$ D = D_c ( \Id + R D_c )^{-1} $ to obtain a local $ D $ which reproduces
correct chiral anomaly on a finite lattice for smooth background
gauge fields.
Therefore, we must explicitly check whether the constraint (e) is satisfied
or not, while the constraint (d) is usually fulfilled except for
bizarre cases.

In order to have $ D_c $ satisfy the constraint (a), we first try
\bea
\label{eq:b}
b = w^{-1} \sum_\mu \sigma_\mu t_\mu
\eea
where 
\bea
\label{eq:t}
t_\mu (x,y) = \frac{1}{2} [   U_{\mu}(x) \delta_{x+\hat\mu,y}
                       - U_{\mu}^{\dagger}(y) \delta_{x-\hat\mu,y} ] \ ,
\eea
\beq
\sigma_\mu \sigma_\nu^{\dagger} + \sigma_{\nu} \sigma_\mu^{\dagger} =
2 \delta_{\mu \nu} \ ,
\eeq
\bea
\gamma_\mu &=& \left( \begin{array}{cc}
                            0                &  \sigma_\mu    \\
                    \sigma_\mu^{\dagger}     &       0
                    \end{array}  \right)
\eea
and $ w $ is a non-singular hermitian operator which
is trivial in the Dirac space and goes to a constant in the classical
continuum limit. Then Eqs. (\ref{eq:DL}) and (\ref{eq:DR}) become
\bea
\label{eq:DLw}
D_L &=& - ( \sigma \cdot t )^{-1}
          \left[ \sqrt{ w^2 + w \ t^2 \ w^{-1} } - w  \right] \\
\label{eq:DRw}
D_R &=& \left[ \sqrt{ w^2 + w \ t^2 \ w^{-1} } + w \right]^{-1}
           ( \sigma \cdot t )
\eea
where
\bea
\label{eq:sigma_t}
\sigma \cdot t &=& \sum_\mu \sigma_\mu t_\mu   \\
\label{eq:t2}
t^2 &=& - ( \sigma \cdot t ) ( \sigma^{\dagger} \cdot t )
\eea
However, $ t_\mu $ suffers from species doublings. Now our task
is to construct a hermitian operator $ w $ such that the doubled
modes are completely decoupled from $ D_c $ or the fermion
propagator $ D_c^{-1} $, even at finite lattice spacing.
The fermion propagator $ D_c^{-1} $ is
\bea
\label{eq:Dci}
D_c^{-1}
\equiv  \left[ \begin{array}{cc}
                      0                 &   D_L^{-1}  \\
                     D_R^{-1}           &   0
                   \end{array}                           \right]
=  \left[ \begin{array}{cc}
                      0                 &  -(\Id + h_1 )^{-1} h_2  \\
       h_2^{-1} (\Id - h_1)             &   0
                   \end{array}                           \right]
\eea
The left-handed free fermion propagator, $ D_L^{-1} $,
in the momentum space is
\bea
\label{eq:DLi}
D_L^{-1} =
- \left[ \sqrt{ w^2 + t^2 } + w \right] \frac{ \sigma_\mu t_\mu }{t^2}
\eea
where $ t_\mu = i \sin( p_\mu a ) $ and $ t^2 = \sum_\mu \sin^2 ( p_\mu a ) $.
The second factor in (\ref{eq:DLi}) is exactly the naive fermion propagator
which suffers from species doubling at the $ 2^d - 1 $ corners of the
Brillouin zone (BZ),
i.e., $ \otimes_{\mu} \{ 0, \pi/a \} \backslash \{ p=0 \} $.
Now we want to construct a hermitian operator $ w $
such that $ w > 0 $ for the primary mode at $ p=0 $ but $ w < 0 $ at
the corners of the BZ, then all doubled modes are decoupled completely
due to the {\it vanishing} of the first factor in (\ref{eq:DLi}).
Explicitly, we want to construct a hermitian $ w $ such that
in the free fermion limit, it satisfies the following condition,
\bea
\label{eq:wp}
w(p) = \left\{  \begin{array}{ll}
  > 0   &  \mbox{ for the primary mode at $ p = 0 $  }     \\
  < 0   &  \mbox{ for the doubled modes at
   $ p \in \otimes_{\mu} \{ 0, \pi/a \} \backslash \{ p=0 \} $.   }  \\
                    \end{array}
              \right.
\eea
It is obvious that $ w(p) $ cannot be a simple and smooth function of
$ \sin( p_\mu a ) $ since it vanishes at all corners of the BZ as well
as at the origin $ p = 0 $.
The next simplest possibility is $ \sin( p_\mu a / 2) $ which vanishes at
the origin $ p = 0 $ but not at the corners of BZ. Therefore the candidate
\bea
\label{eq:wp1}
w(p) = c - \sum_\mu  \sin( p_\mu a / 2), \hspace{4mm} c \in ( 0, 1)
\eea
would satisfy the condition (\ref{eq:wp}).
But (\ref{eq:wp1}) can only be realized by the
symmetric difference operator on a lattice with lattice spacing $ a/2 $.
If we wish to retain the simple lattice with lattice spacing $ a $,
we need to modify the form of $ \sin( p_\mu a / 2) $ in (\ref{eq:wp1}).
The simplest modification one can perform is to take the square of
$ \sin( p_\mu a / 2) $ since
$ \sin^2 ( p_\mu a / 2) = [ 1 - \cos ( p_\mu a ) ] / 2 $ which
can be implemented by nearest neighbor difference operators on the
lattice with lattice spacing $ a $. So, we modify (\ref{eq:wp1})
to
\bea
\label{eq:wpc}
w(p) = c - \sum_\mu  \sin^2 ( p_\mu a / 2), \hspace{4mm} c \in ( 0, 1 )
\eea
It should be emphasized that the role
of $ w $ in our general solution of $ D_c $ is quite different from the
Wilson term in the Wilson-Dirac operator \cite{wilson75}. In our
general solution of $ D_c $, the chiral symmetry is always preserved,
and {\it the role of $ w $ is to suppress the doubled modes completely
at finite lattice spacing}, while in the Wilson-Dirac operator,
the Wilson term breaks the chiral symmetry explicitly and gives a mass
of order $ a^{-1} $ to the doubled modes such that they can be
decoupled in the continuum limit ( $ a \to 0 $ ).
After the gauge links are restored, $ w $ in the position space becomes
\bea
\label{eq:wxy}
w(x,y) = c \ \delta_{x,y} - \frac{1}{4} \sum_\mu \left[ 2 \delta_{x,y}
                       - U_{\mu}(x) \delta_{x+\hat\mu,y}
                       - U_{\mu}^{\dagger}(y) \delta_{x-\hat\mu,y} \right]
\eea
This is one of the simplest
solution of $ w $ satisfying the requirement (\ref{eq:wp}) in the free
field limit. It is plausible that there exists other solutions to
(\ref{eq:wp}). Here we do not intend to discuss all possible solutions
of $ w $, but only to formulate the necessary condition (\ref{eq:wp})
for $ w $ to decouple all doubled modes at finite lattice spacing.

Since the $ D_c $ [ Eqs. (\ref{eq:DLw})-(\ref{eq:DRw})
with $ t_\mu $ and $ w $ defined in (\ref{eq:t}) and (\ref{eq:wxy}) ]
in the free fermion limit is free of species doubling,
and in the momentum space behaves like $ i \gamma_\mu p_\mu $ as $ p \to 0 $,
the perturbation calculation in ref. \cite{twc99:1} showed that
$ D = D_c ( \Id + r D_c )^{-1} $ gives the correct chiral anomaly.
However, the topological characteristics \cite{twc98:9a} of a lattice
Dirac operator is an intrinsically non-perturbative quantity which
in general cannot be revealed by perturbation calculations.
Therefore further non-perturbative ( e.g., numerical ) studies are
needed before one can understand the topological characteristics
of this $ D_c $. The main objective of this paper is to cast the
general solution of $ D_c $ in the form of (\ref{eq:DL})-(\ref{eq:DR}),
and to demomstrate a viable construction of the operator $ b $.

For any $ D_c $ satisfying criteria (a)-(e), we can use the
topologically invariant transformation $ D = D_c ( \Id + R D_c )^{-1} $ to
obtain a class of local and well defined Dirac fermion operators
which not only preserve the physics of $ D_c $ but also
restrict the quantum corrections to behave properly.
The fermion propagator and chiral condensate of $ D $ have been
investigated in ref. \cite{twc98:10b}.

In summary, we have constructed a class of general solution of the
chirally symmetric $ D_c $ in terms of the hermitian operator $ w $ which
fulfils the requirement (\ref{eq:wp}) in the free field limit.
An example of $ w $ is given in (\ref{eq:wxy}).
One of the salient features of the present
formulation is that the chiral fermion operator $ D_L $ ( $ D_R $ )
is obtained explicitly. This may provide a starting point for a
formulation of chiral gauge theories on a finite lattice.

\bigskip
\bigskip

\flushpar
{\bf Acknowledgement }
\bigskip

\noindent
This work was supported by the National Science Council, R.O.C.
under the grant number NSC89-2112-M002-017.

%\bigskip
%\bigskip

\vfill\eject

\vfill\eject

\end{document}